\documentstyle[twocolumn,epsf]{article}
\begin{document}
\title{ Energetics of Forced Thermal Ratchet 
}
\author{Hideki Kamegawa, Tsuyoshi Hondou \& Fumiko Takagi \\ \\
  Department of Physics,
 Tohoku University  
 \\  Sendai 980-8578, Japan}
\maketitle

{\bf 
 Molecular motors are known to have the high efficiency of energy 
transformation 
in the presence of thermal fluctuation\cite{Woledge}.
 Motivated by the surprising fact, 
 recent studies of thermal ratchet models\cite{Feynman}
 are showing how and when
work should be extracted from 
non-equilibrium fluctuations\cite{A1,A2,Vale,Magnasco,Ajdari,Doering,HondouPRL,RMP}.
 One of the important finding was brought by Magnasco\cite{Magnasco} 
 where he studied the temperature dependence on the fluctuation-induced
current in a ratchet (multistable) system and showed that
the current 
 can generically be maximized in  a finite temperature. 
 The interesting finding has been interpreted that thermal fluctuation
is {\em not} harmful for the fluctuation-induced work and even 
facilitates its efficiency. We show, however, this interpretation  
turns out to be incorrect as soon as we go into the realm of 
the energetics\cite{Sekimoto}: the efficiency of energy transformation 
is not maximized at finite temperature, even in the
same system\cite{Magnasco} that Magnasco 
considered.
The maximum efficiency is realized in the absence of thermal fluctuation.
The result presents an open problem whether thermal fluctuation 
could facilitate the efficiency of energetic transformation from
force-fluctuation into work.
}

\begin{figure}[b]

\caption{
 Schematic illustration of the potential,  $V(x)=V_0(x)+V_L(x)$.
$V_0(x)$ is a piecewise linear and periodic potential.
$V_L(x)$ is a potential due to the load, $V_L(x)=lx$.
The period of the potential is $\lambda = \lambda _1 + \lambda _2 $,
 and $\Delta
 =\lambda _1 - \lambda _2 (> 0)$ is a symmetry breaking amplitude.
For a finite work against load, we assume the condition $l \le A$ throughout 
this paper.
}
\end{figure}

Let us consider a forced ratchet system subject to an external load:
     \begin{equation}
        \frac{d x}{d t} = -\frac{\partial V_{0}(x)}{\partial x} + \xi (t) +F(t) 
                  -\frac{\partial V_{L}(x)}{\partial x},
\label{eq:1}
\end{equation}
where $x$ represents the state of the ratchet, 
$V_{0}(x)$ is a periodic potential, $\xi (t)$ is a thermal noise satisfying
        $<\xi (t)\xi (t')> = 2kT\delta (t-t')$,
"$< \cdot >$" is an operator
of ensemble average,
$F(t)$ is an external fluctuation,
 $F(t + \tau) = F (t)$, $\int_{0}^{\tau} dt F(t) =0$, 
and $V_{L}$ is a potential 
due to the load,
  $\frac{\partial V_{L}}{\partial x} =l > 0 $.
The geometry of the potential, $V(x) = V_{0}(x)  + V_{L}(x)$, is displayed
in  Fig. 1. 
 The ratchet system transforms the external fluctuation into
work (see, for review, Ref\cite{RMP}).
 The model\cite{Magnasco} Magnasco discussed is a special case of the present
system,  where the external
load is omitted.
 In general,
  Fokker-Planck equation\cite{Gardiner} of the system is written:
     \begin{equation}
\begin{array}{rcl}
&	\frac{\partial P(x,t)}{\partial t} + 
\frac{\partial J(x,t)}{\partial 
x} = 0, & \\
&	J = -kT \frac{\partial P(x,t)}{\partial x} + \{
 - \frac{\partial V_{0}(x)}{\partial x} 
 +F(t)-l\} P(x,t), &
\end{array}
     \end{equation}
where
 $P(x,t)$ is a probability density and $J(x,t)$ is a probability
current.
If $F(t)$ changes slowly enough,  $P(x,t)$ could be treated as quasi-static.
 In such situation, $J$ can be obtained in an analytical form. 
For slowly changing fluctuation  $F(t)$ of square wave\cite{Magnasco} 
of amplitude $A$, we analytically obtain
 an average current over the period of the 
fluctuation, 
     \begin{eqnarray}
	J_{sqr} &=& \frac{1}{2} [ J(A) + J(-A) ].
\label{eq:sqr}
     \end{eqnarray}

It is reported\cite{Magnasco} in the operation of the ratchet that 
 "there is a region of operating regime where
the efficiency is optimized at finite temperature."
The result has been interpreted that the operation of the  forced thermal
 ratchet is helped by  thermal fluctuation.
This discovery has been followed and confirmed by many literatures (see
 the references in Ref.\cite{RMP})
 with various situations.
We first confirm the previous report and then analyze it energetically. 
We can distinguish the behavior of the current $J_{sqr}$ on the temperature
into three regimes, ($a$)
	$ A< \frac{Q}{\lambda _1} +l  < \frac{Q}{\lambda _2} -l$, 
	($b$) $\frac{Q}{\lambda _1} +l  < A < \frac{Q}{\lambda _2} - l$, and
        ($c$) $\frac{Q}{\lambda _1} +l <  \frac{Q}{\lambda _2} -l  < A$;
 whereas
the distinction is not explicitly described in the paper\cite{Magnasco}.
In regimes ($a$) and ($b$), 
$J_{sqr}$ is certainly maximized at finite temperature (Fig. 2$a$, $b$).
In regime 
($c$), $J_{sqr}$ is a monotonically decreasing function of the
 temperature (Fig. 2$c$).

\begin{figure}[b]

\caption{
Plot of the current $J_{sqr}$ as a function of $kT/Q$.
The first regime (a),  $A < \frac{Q}{\lambda _1}+l < \frac{Q}{\lambda _2}-l$ 
    ($\lambda =1.0$,\ $\Delta =1.0$,\ $l=0.01$,\ $A=1.0$); 
the second regime (b),  $\frac{Q}{\lambda _1}+l < A < \frac{Q}{\lambda _2}-l$ 
   ($\lambda =1.0$,\ $\Delta =1.0$,\ $l=0.01$,\ $A=1.2$); 
and the third regime (c),  $\frac{Q}{\lambda _1}+l < \frac{Q}{\lambda _2}-l < 
A$  
   ($\lambda =1.0$,\ $\Delta =0.6$,\ $l=0.01$,\ $A=6.0$).
Regimes (a), (b) and (c) correspond to the low, 
moderate and high amplitude forcing in the description of Magnasco[6] 
respectively.
In regimes (a) and (b),
 increasing temperature results first in a rise and then a fall in the current.
}
\end{figure}

We have to notice at this stage that the fluctuation-induced current $J$ is not 
an energetic quantity and therefore $J$ is only the mimic of the 
{\em energetic} efficiency.  
The lack of discussion of the forced ratchet system 
by this {\em real} efficiency is attributed to 
the lack of construction of energetics of the systems described by
Langevin or equivalently by Fokker-Planck equations.
Recently, an energetics of these systems was constructed
by Sekimoto\cite{Sekimoto}.
Therefore, we will go into the realm of the energetics of the forced 
thermal ratchet, and analyze the {\em real} efficiency.

According to  the energetics\cite{Sekimoto}, 
the input energy $R$ per unit time from external force to the ratchet
and the work $W$ per unit time 
that the ratchet system extracts from the fluctuation into 
the work are written respectively:
\begin{equation}
	R[F(t)]  =  \frac{1}{t_{f}-t_{i}}\int_{x=x(t_i)}^{x=x(t_f)} F(t)
 dx(t), 
\end{equation}
   \begin{equation}
        W = \frac{1}{t_f - t_i} \int_{x=x(t_i)}^{x=x(t_f)} dV(x(t)).
     \end{equation}
For the square wave,
they yield:
\[
	<R_{sqr}> = \frac{1}{2} [<R(A)>+<R(-A)>] \nonumber
\]
\begin{equation}
		 = \frac{1}{2} A[J(A) - J(-A)],
     \end{equation}
     \begin{equation}
	<W_{sqr}> = \frac{1}{2} l[J(A)+J(-A)].
     \end{equation}
Therefore, we obtain the efficiency of the energy transformation $\eta$:
     \begin{equation}
	\eta = \frac{<W_{sqr}>}{<R_{sqr}>} \nonumber \\
	     = \frac{l[J(A)+J(-A)]}{A[J(A)-J(-A)]}.
      \label{eff}
     \end{equation}
Because $\frac{J(-A)}{J(A)} < 0$, Eq. (\ref{eff}) is rewritten,
\begin{equation}
 \eta = \frac{l}{A} \left\{ 1 -
             \frac{2 |\frac{J(-A)}{J(A)}|}{1+|\frac{J(-A)}{J(A)}| } \right\} .
\label{abs}
\end{equation}
 In the limit $|\frac{J(-A)}{J(A)}| \rightarrow 0$, 
 the maximum efficiency of the energy transformation for given load $l$
and force amplitude $A$ is realized: 
$\eta_{max} = \frac{l}{A}$. 

\begin{figure}[b]

\caption{
Plot of the efficiency $\eta $ as a function of $kT/Q$.
In each regime (a), (b) and (c), the condition is the same as in Fig.2. 
In all the regimes, increasing temperature results in decreasing the efficiency.
}
\end{figure}

 In Eq. (\ref{abs}), we can discuss the effect of thermal fluctuation on 
the {\em energetic} efficiency of the forced thermal ratchet. As
demonstrated in Fig. 3, it is proved that the efficiency is a decreasing 
function of temperature in all the three regimes:
Because,  $|\frac{J(-A)}{J(A)}|$ is monotonically increasing 
function of the temperature as found in Fig. 4,
the efficiency $\eta$ is 
 a decreasing function of the temperature.
This certainly shows that the presence of thermal fluctuation {\em does not}
help efficient energy transformation by the ratchet, which is in contrast to 
the previous interpretation that thermal fluctuation could facilitates 
the efficiency.

\begin{figure}[b]

\caption{
Plot of currents $J(A)$, $J(-A)$ and $\left| \frac{J(-A)}{J(A)} \right| $.
The condition is the same as in the second regime (b) in Fig. 2.
$|J(-A)|$ increases slower than  $|J(A)|$ when $kT$ increases
 from zero temperature.
This difference is attributed to the symmetry breaking of the potential 
as illustrated in Fig. 1.
}
\end{figure}

 We can learn here that the efficiency should be discussed energetically:
 The condition of maximum current does not correspond to that of the
maximum efficiency. The difference is attributed to the observation that
the efficiency is a ratio of the extracted work $W$ to the
 consumed energy $R$.
 The extracted work  $W$ is surely proportional to the current
  $J_{sqr} = \frac{1}{2} [ J(A) + J(-A) ]$ (Eq. \ref{eq:sqr}).
However, the consumed energy is not a constant but 
 varies sensitively according to the
condition. Therefore the efficiency $\eta$ is not simply proportional to
the induced current $J$.
The important problem was left for future study whether the existence
of thermal fluctuation could facilitate the efficiency of energy
transformation in 
more general forced ratchets. 

Finally, we mention that
the complementarity relation\cite{Sasa} of the ratchet system.
 We found that the maximum efficiency, $\eta_{max} = 1$, can be realized 
if all of the 
the following conditions are satisfied, $A \rightarrow Q/\lambda_{1}+l+0$ and 
 $Q \rightarrow +0$ and $T \rightarrow 0$.
 In this limit, the speed of energy transformation goes to zero.
That is to say, this maximum efficiency of the forced ratchet is 
realized in quasistatic process.
As we increase the velocity of this engine, the efficiency is decreased.
 The result also emphasizes the importance of time scales of the operation
of the ratchet as J\"ulicher et al. pointed out\cite{RMP}. 
 Detailed analysis of the loss of the efficiency may be
 analyzed by the formal theory  of
complementarity relation\cite{Sasa} between the time lapse 
of thermodynamic process 
and the irreversible heat. 

\hspace{10mm}

 {\bf Acknowledgements.}
 We would like to thank
K. Sekimoto, S. Sasa, T. Fujieda and T. Tsuzuki for helpful comments
and discussions.
This work is supported
by the Japanese Grant-in-Aid for Science Research Fund from the Ministry
of Education, Science and Culture.
\\
\\
Correspondence and requests for materials should be addressed to T.H.  \\
(e-mail: hondou@cmpt01.phys.tohoku.ac.jp).

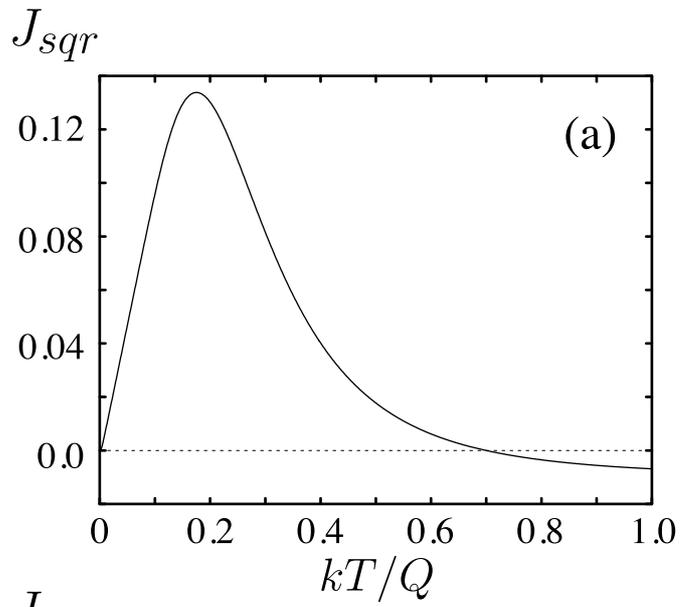
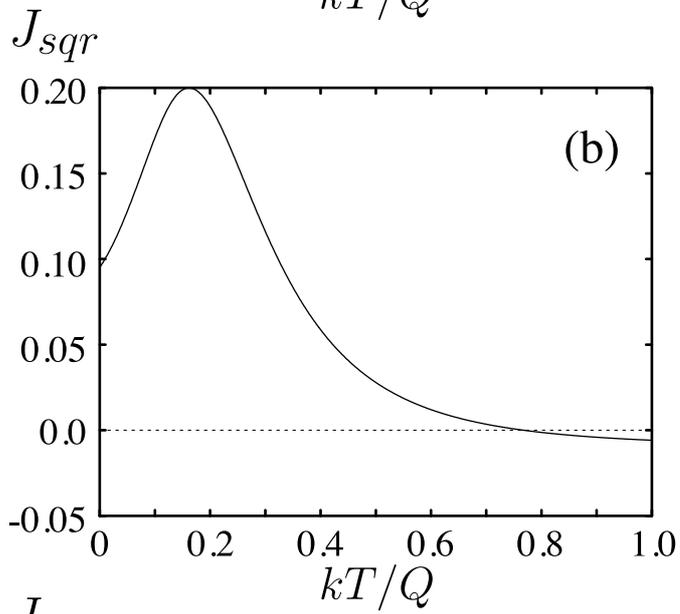
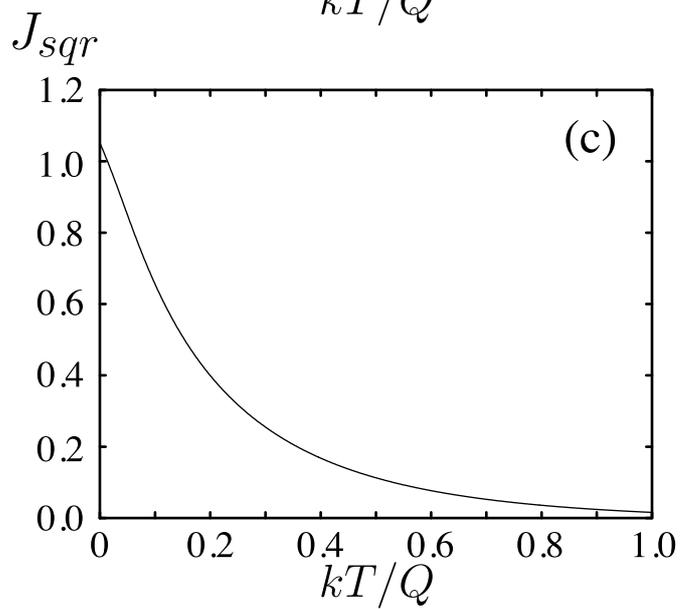
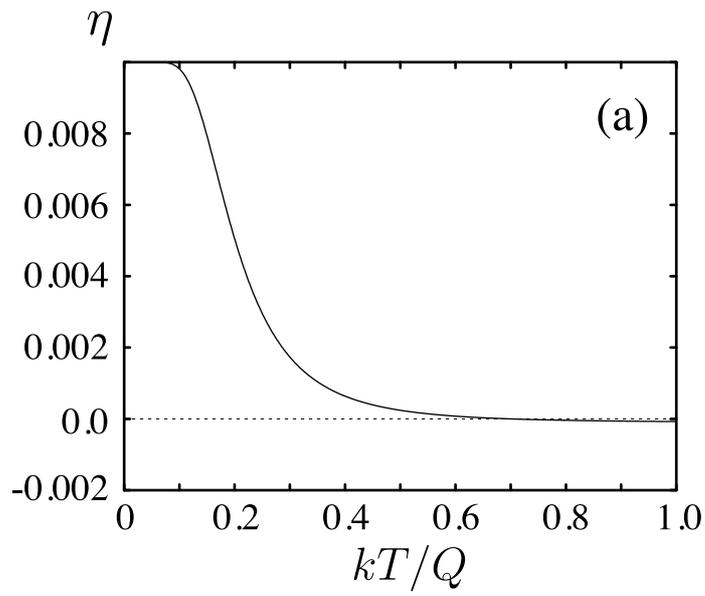
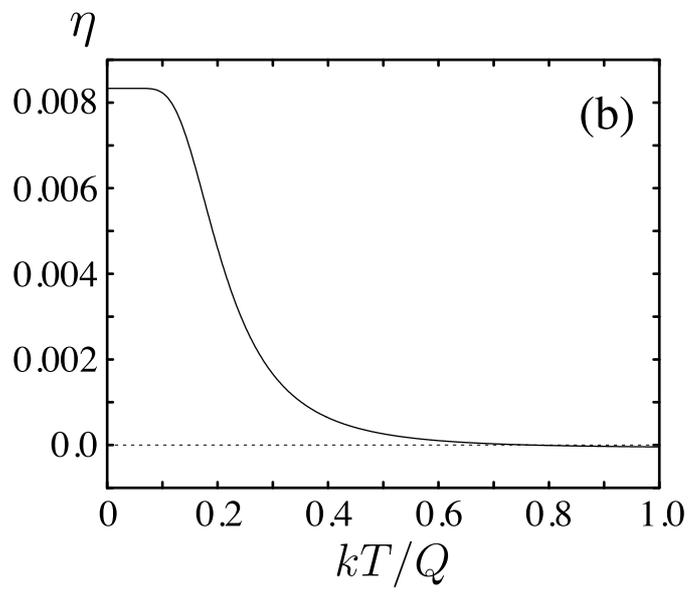
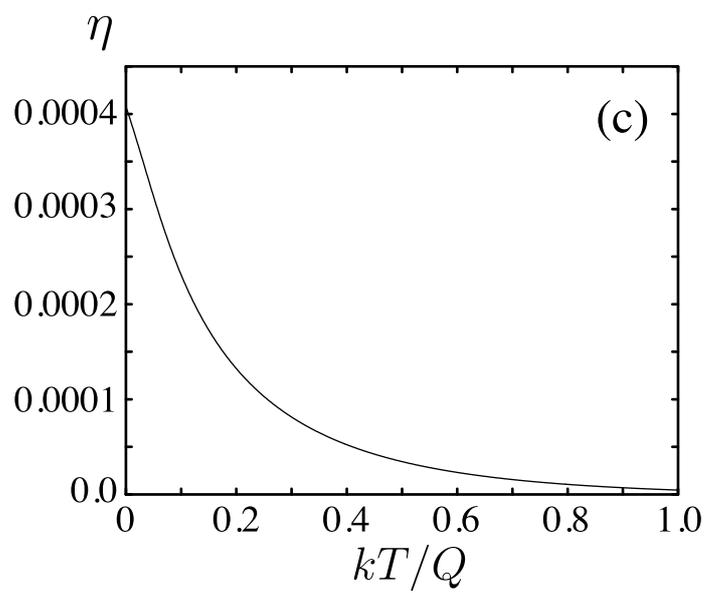
\end{document}